\documentclass[journal]{IEEEtran}

\usepackage{amsmath,epsfig}

\usepackage{graphicx}
\usepackage{color}
\usepackage{placeins}
\usepackage{float}
\usepackage{tabularx,colortbl}
\usepackage{epstopdf}
\usepackage{amsmath}
\usepackage{amssymb}
\usepackage{caption}
\usepackage{subcaption}
\usepackage{algorithm}
\usepackage{algorithmic}
\usepackage{multirow}
\usepackage{url}

\newtheorem{Lem}{Theorem}

\begin{document}

\title{Power Control for Wireless VBR Video Streaming: From Optimization to Reinforcement Learning}

\author{Chuang Ye, M. Cenk Gursoy, and Senem Velipasalar
\thanks{The authors are with the Department of Electrical
Engineering and Computer Science, Syracuse University, Syracuse, NY, 13244
(e-mail: chye@syr.edu, mcgursoy@syr.edu, svelipas@syr.edu).}}

\maketitle

\thispagestyle{empty}
\begin{abstract}
In this paper, we investigate the problem of power control for streaming variable bit rate (VBR) videos over wireless links. A system model involving a transmitter (e.g., a base station) that sends VBR video data to a receiver (e.g., a mobile user) equipped with a playout buffer is adopted, as used in dynamic adaptive streaming video applications. In this setting, we analyze power control policies considering the following two objectives: 1) the minimization of the transmit power consumption, and 2) the minimization of the transmission completion time of the communication session. In order to play the video without interruptions, the power control policy should also satisfy the requirement that the VBR video data is delivered to the mobile user without causing playout buffer underflow or overflows. A directional water-filling algorithm, which provides a simple and concise interpretation of the necessary optimality conditions, is identified as the optimal offline policy. Following this, two online policies are proposed for power control based on channel side information (CSI) prediction within a short time window. Dynamic programming is employed to implement the optimal offline and the initial online power control policies that minimize the transmit power consumption in the communication session. Subsequently, reinforcement learning (RL) based approach is employed for the second online power control policy. Via simulation results, we show that the optimal offline power control policy that minimizes the overall power consumption leads to substantial energy savings compared to the strategy of minimizing the time duration of video streaming. We also demonstrate that the RL algorithm performs better than the dynamic programming based online grouped water-filling (GWF) strategy unless the channel is highly correlated.
\end{abstract}
\section{Introduction}
\label{sec:intro}

Multimedia applications such as video telephony, teleconferencing, and video streaming have started becoming predominant in data transmission over wireless networks. For instance, as reported in \cite{Cisco}, mobile video traffic exceeded 50\% of total mobile data traffic for the first time in 2012, and grew to 60\% in 2016, and more than three-fourths of the global mobile data traffic is expected to be video traffic by 2021. These applications are very sensitive to sudden degradations in channel quality or outage, which may lead to video packet loss and play interruption at the receiver side. One approach to address this problem is to adapt the transmission power according to the variations of the channel conditions. For instance, in order to maintain high-quality dynamic adaptive video streaming, a higher transmit power can be used when the channel conditions are poor and the power levels can be reduced when the channel conditions improve. However, this may lead to a significant increase in the energy consumption in the transmission session. On the other hand, reducing the energy consumption of data delivery is becoming an increasingly important challenge in order to utilize scarce energy resources, reduce costs, and sustain green operation. If the demand profile and channel conditions are known beforehand, this information can be used to cache the data in the receiver's buffer in advance when channel conditions are good and the buffered data can be played to maintain continuous video streaming even when the channel conditions are poor. Such pre-downloading of data can enable us to balance the power levels in different time slots, leading to substantial savings in energy consumption.

Motivated by the above considerations, wireless video streaming has been addressed in several recent studies. For instance, scheduling algorithms to transmit multiple video streams from a base station (BS) to mobile clients were investigated in \cite{seetharma}. With the proposed algorithms, the vulnerability to stalling was reduced by allocating slots to videos in a way that maximizes the minimum playout lead across all videos within an epoch-by-epoch framework. Authors in \cite{yhuang} proposed algorithms to find the optimal transmit powers for the base stations with the goal of maximizing the sum transmission rate such that the variable bit rate (VBR) video data can be delivered to the mobile users without causing playout buffer underflow or overflows. A deterministic model for VBR video traffic that considers video frame sizes and playout buffers at the mobile users was adopted. In \cite{habou}, the authors investigated an energy-efficient video downlink transmission by predicting the download rate at the receiver. In \cite{chuang-wcnc18}, we studied power control and mode selection for VBR video streaming in D2D networks in the presence of potential interference, and showed that video delivery with power control and mode selection leads to improved performance. Power control is determined by judiciously considering all possible scenarios and checking the constraints. In \cite{chuang-TCOM18}, we employed effective capacity as the throughput metric and analyzed quality-driven resource allocation for full-duplex delay-constrained wireless video transmissions.

In \cite{JWu}, the authors developed an analytical framework to characterize the energy-distortion relationship in multipath video wireless transmissions over heterogenous networks. The minimization of energy consumption was achieved by optimally allocating the video flow rate under the target video quality constraint. The authors in \cite{Jwu1} proposed a bandwidth aggregation framework, which integrates energy-minimized rate adaption, delay-constrained unequal protection and quality-aware packet distribution, to enable energy-minimized video quality-guaranteed streaming to multihomed devices within the imposed deadline. The authors in \cite{Ahmad} developed a distributed joint power control and rate adaptation framework for video streaming in multi-node wireless networks within a time-varying interference environment. The optimal power allocation is conducted in order to achieve a certain target signal-to-interference-plus-noise ratio, such that the difference between the arrival and the departure rates at the queues is very small. Rate adaption was performed according to the video quality demand, channel conditions and a given fairness criterion.

In \cite{Stockhammer}, the author enabled decoding of each video unit before exceeding the playout deadline. Therefore, the successful video sequence presentation can be guaranteed even if the media rate does not match the constant or VBR channel rate. It also showed that the separation between a delay jitter buffer and a decoder buffer is suboptimal for VBR video transmission over wireless channels. The authors in \cite{Chatziperis} and \cite{Rango} investigated efficient admission control schemes for VBR videos over wireless networks in terms of bandwidth and QoS requirements based on Discrete
Autoregressive (DAR (1)) model and the statistical multiplexing of VBR traffic.

The relationship between the transmission rate and distortion of received video sequences also plays an important role in video wireless transmission and have been addressed in e.g., \cite{Zhihai1} -- \cite{Zhihai4}. In \cite{Zhihai1}, the authors studied how the resource constraints in wireless video communication could be incorporated into the rate-distortion (R-D) analysis, and developed a resource-distortion analysis framework. The authors in \cite{Zhihai2} and \cite{Zhihai3} proposed an analytic power-rate-distortion (P-R-D) model to characterize the relationship between the power consumption of a video encoder and its rate-distortion performance. Based on this model, the optimum power allocation between video encoding and wireless transmission under energy constraints was studied. This P-R-D model has been used by portable video communication devices in \cite{Zhihai4} with the goal of minimizing the energy consumption.

Finally, we note that while not directly addressing multimedia transmissions, several recent studies on energy harvesting exhibit a certain level of parallelism to the problems considered in this paper. For instance, optimal packet scheduling problem in a single-user energy harvesting wireless communication system was studied in \cite{jyang}. The time by which all packets are delivered was minimized by adaptively changing the transmission rate according to the traffic load and available energy. The problem of online packet scheduling to minimize the required conventional grid energy for transmitting a fixed number of packets given a common deadline was considered in \cite{adeshmukh}. The proposed algorithm aims to finish the transmission of each packet assuming that all future packets are going to arrive at equal time intervals within the left-over time. The authors in \cite{dshaviv} considered online power control with the goal of maximizing the long-term average throughput in an energy harvesting system with random independent and identically distributed (i.i.d.) energy arrivals and a finite battery for data transmission. A simple online power control policy was proved to be universally near-optimal for all parameter values.

In this paper, we consider the problem of dynamic adaptive streaming of VBR videos (for instance, in applications such as YouTube and Netflix) over multiple subchannels in a wireless link. Note that VBR video has stable video quality within the frames at the cost of large variations in the frame size or bit rate, whereas constant-bit-rate (CBR) video has a stable bit rate but the visual qualities of the frames vary significantly. Within this, we consider a traffic model for stored VBR video, taking into account the frame size, frame rate, and playout buffers \cite{sen}, \cite{liang}. We exploit power control over multiple subchannels at the transmitter with the goal of minimizing the overall energy consumption without underflow and overflows. More specifically, our contributions can be listed as follows:
\begin{enumerate}
\item We formulate optimization problems to minimize the overall power consumption and characterize the optimal allocation of power across subchannels and over time subject to buffer overflow and underflow constraints.

\item We identify the directional water-filling power control as the optimal offline policy and develop a dynamic programming based novel power control algorithm that utilizes the key properties of the optimal water-filling policy (e.g., on the optimal power characterizations and time-varying water levels).

\item In addition to the minimization of the power consumption, we address the minimization of the time duration of video streaming, and similarly characterize the optimal power control strategies and develop an offline algorithm.

\item We design two novel and efficient online power control schemes under the practically appealing assumption that only the current channel fading state is known and future states are predicted. While the first scheme applies directional water-filling approach with predicted channel states within a certain time window, the second online policy is based on reinforcement learning and incorporates buffer overflow and underflow constraints, channel prediction, and water-filling type power allocation strategies in the selection of feature functions.
\end{enumerate}

The remainder of this paper is organized as follows: The system model is presented in Section \ref{sec:System_Model}. Optimization problems are formulated and the optimal offline policies are identified in Section \ref{sec:offlinePolicy}.  Efficient online policies are determined in Section \ref{sec:onlinePolicy}.  Numerical results are presented and discussed in Section \ref{sec:Result}. Finally, we conclude the paper in Section \ref{sec:Conclusion}.

\section{System Model} \label{sec:System_Model}
\begin{figure}
\centering
\includegraphics[width=0.5\textwidth]{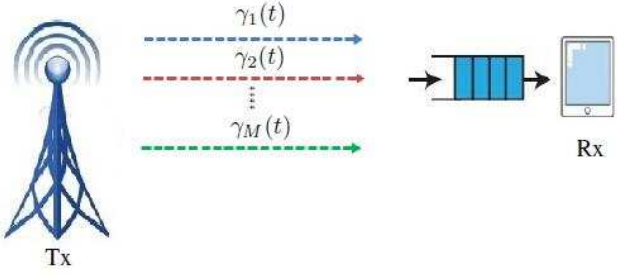}
\caption{\small{System model for VBR video streaming over a wireless link with multiple subchannels. }}\label{fig:System_Model}
\end{figure}

We consider video streaming over a wireless fading link with multiple subchannels as shown in Fig. \ref{fig:System_Model}. The arriving data is stored in a playout buffer at the receiver ($\text{Rx}$). There are $M$ orthogonal subchannels between the transmitter ($\text{Tx}$) and $\text{Rx}$ with bandwidth $B_c$ for each subchannel, and the total bandwidth is $B = M B_c$. We assume that each channel experiences block-flat fading during each time slot $t$. Thus, the capacity of the $i^{\text{th}}$ subchannel in time slot $t$ is
\begin{equation}
C_i(t) = B_c \log \left(1 + \frac{P_i(t)\gamma_i(t)}{N_0 B_c}\right), \label{eq:capacity_sub}
\end{equation}
where $P_i(t)$ and $\gamma_i(t)$ are the transmission power and ergodic and stationary fading power in the $i^{\text{th}}$ subchannel in time slot $t$, respectively. $N_0$ is the power spectral density of the background Gaussian noise.
Therefore, the total throughput over all the subchannels in time slot $t$ is $C(t) = \sum_{i=1}^{M} C_i(t)$.

\begin{figure}
\centering
\includegraphics[width=0.5\textwidth]{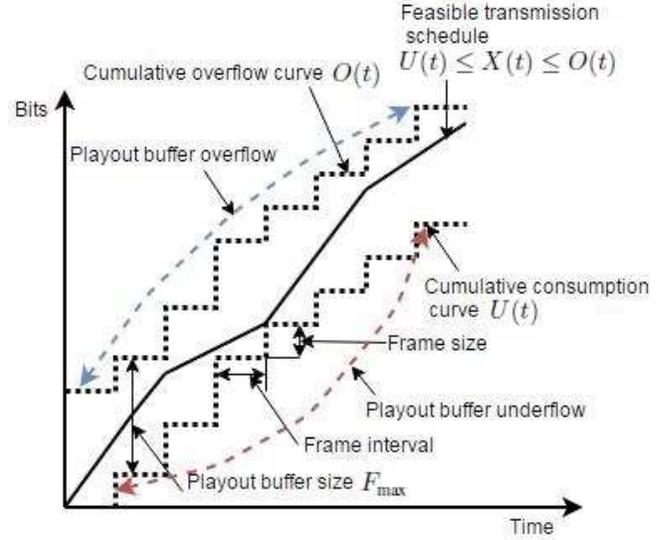}
\caption{\small{Feasible and infeasible transmission schedules for video.}}\label{fig:Buffer}
\end{figure}

Let $F(t)$ be the video consumption/frame size at the $\text{Rx}$, representing the amount of data played by the video player in time slot $t$. We assume that the video has $T$ frames, and due to the limited storage, the playout buffer size at the $\text{Rx}$ is $F_{\text{max}}$. Let $U(t)$ represent the cumulative consumption curve at time $t$, representing the cumulative amount of bits consumed by the $\text{Rx}$. The remaining data in the buffer at time $t$, which is denoted by $D(t)$, should not exceed the buffer storage size. In the meantime, in order to play the video without any interruption at the $\text{Rx}$, $D(t)$ should not be less than the frame size required at time $t$, $F(t)$. Therefore, the constraints for the remaining data in the buffer at time $t$ are formulated as follows:
\begin{align}
& D(t) \leq F_{\text{max}}, \quad \quad  \quad 0 \leq t \leq T,    \label{eq:overflow1} \\
& D(t) \geq F(t), \quad \quad \quad 0 \leq t \leq T,    \label{eq:underflow1}
\end{align}
where we assume $F(0) = 0$.
The remaining data $D(t)$ depends on the arrival data at time $t$ and consumed data at time $t-1$, and thus the relation among remaining data and arrival data and consumed data is expressed as follows:
\begin{align}
D(t) = D(t-1) - F(t-1) + C(t)\tau, \quad 1 \leq t \leq T,  \label{eq:data_relation1}
\end{align}
where $\tau$ is the duration of one time slot and we assume $D(0) = 0$. After some straightforward manipulations, (\ref{eq:data_relation1}) can be rewritten as
\begin{align}
D(t) & = \sum_{i=1}^{t}C(i)\tau-\sum_{i=1}^{t-1}F(i) \nonumber \\
    & = X(t) - U(t-1), \quad 1 \leq t \leq T,   \label{eq:data_relation2}
\end{align}
where $X(t) = \sum_{i=1}^{t}C(i)\tau$ denotes the amount of cumulative arrival data at time $t$. Let $O(t)$ denote the cumulative overflow curve, representing the maximum cumulative amount of bits that does not violate the buffer length constraint. Hence, $O(t)$ and $U(t)$ can be expressed as
\begin{align}
& O(t) = \sum_{i = 0}^{t-1} F(i) + F_{\text{max}}, \quad 1 \leq t \leq T,\label{eq:overflow_form}\\
& U(t) = \sum_{i = 1}^{t} F(i), \quad \quad \quad \quad 1 \leq t \leq T. \label{eq:underflow_form}
\end{align}
Therefore, constraints in (\ref{eq:overflow1}) and (\ref{eq:underflow1}) can now be rewritten as
\begin{align}
& X(t) \leq O(t), \quad \quad  1 \leq t \leq T, \label{eq:overflow2}\\
& X(t) \geq U(t), \quad \quad  1 \leq t \leq T, \label{eq:underflow2}
\end{align}
and Fig. \ref{fig:Buffer} shows that a feasible transmission schedule will generate a cumulative transmission curve $X(t)$ that lies within $O(t)$ and $U(t)$ in order to play the video without interruptions (i.e., without buffer overflows and underflows).

Finally, we note that while not specifically discussed above, transmission delay is an implicit component in the analysis, and is inversely proportional to the transmission power level. In particular, transmission delay of a video frame can be formulated as the frame size divided by the throughput $C(t)$. Therefore, while controlling the transmission power level across time and multiple subchannels, we also essentially keep in check the transmission delay. Specifically, while too large a transmission delay can lead to buffer underflows and hence video stalls, allocating excessive power levels and sending too many frames with short delays can incur buffer overflows.

\section{Optimal Offline Policies}\label{sec:offlinePolicy}
In this section, we analyze optimal offline policies. As our primary goal, we initially characterize the optimal policy that minimizes the power consumption in wireless video streaming. Subsequently, we will address the minimization of time duration of video streaming in order to provide comparisons with the power minimization policies. In both cases, overflow and underflow constraints will be imposed.

\subsection{Minimizing power consumption}
In this section, the goal is to determine the optimal offline policy that minimizes the overall power consumption under the requirement that $\text{Rx}$ plays the received video without any interruptions and missing frames (i.e., without any overflows and underflows in the playout buffer). Therefore, the optimization problem can be expressed as follows:

\begin{align}
& \min_{\mathbf{P}}  \sum_{j = 1}^{T} \sum_{i = 1}^{M} P_i(j) \nonumber \tag{\textbf{P1}}\label{prb:p1} \\
\text{s.t. } & \sum_{j = 1}^{t} \sum_{i = 1}^{M} C_i(j) \tau \geq \sum_{j = 1}^{t} F(j), \forall t = 1, \ldots, T-1, \label{eq:cond_underflow1}\\
& \sum_{j = 1}^{T} \sum_{i = 1}^{M} C_i(j) \tau  = \sum_{j = 1}^{T} F(j), \label{eq:cond_total1} \\
& \sum_{j = 1}^{t} \sum_{i = 1}^{M} C_i(j) \tau \leq \sum_{j = 1}^{t-1} F(j)+F_{\text{max}}, \forall t = 1, \ldots, T, \label{eq:cond_overflow1}
\end{align}
where $\mathbf{P}$ is an $M\times T$ dimensional power matrix with the component in the $i^{\text{th}}$ row and $j^{\text{th}}$ column $P_i(j)$ denoting the power allocated to the $i^{\text{th}}$ channel at time $j$. (\ref{eq:cond_underflow1}) and (\ref{eq:cond_overflow1}) are the minimum cumulative data requirement and buffer overflow violation constraints, respectively, described in Section \ref{sec:System_Model}. (\ref{eq:cond_total1}) is the constraint that the overall received data should be equal to the size of the transmitted video.

The objective function in Problem (\ref{prb:p1}) is a linear function of $\mathbf{P}$. However, since the constraint (\ref{eq:cond_overflow1}) is a concave function with respect to $\mathbf{P}$, the optimization problem (\ref{prb:p1}) is not in the form of a convex optimization problem.

On the other hand, (\ref{eq:capacity_sub}) can be rewritten as
\begin{equation}
P_i(t) = \left(2^{\frac{C_i(t)}{B_c}}-1\right)\frac{N_0 B_c}{\gamma_i(t)}, \label{eq:power}
\end{equation}
and therefore, the optimization problem (\ref{prb:p1}) can also be reformulated in terms of $C_i(j)$ as
\begin{align}
& \min_{\mathbf{C}}  \sum_{j = 1}^{T} \sum_{i = 1}^{M} \left(2^{\frac{C_i(j)}{B_c}}-1\right)\frac{N_0 B_c}{\gamma_i(j)} \nonumber \tag{\textbf{P2}}\label{prb:p2} \\
\text{s.t. } & \quad \quad \quad (\ref{eq:cond_underflow1}), (\ref{eq:cond_total1}), (\ref{eq:cond_overflow1}).  \nonumber
\end{align}
Above, the objective function of (\ref{prb:p2}) is an increasing convex function of $\mathbf{C}$, and constraints (\ref{eq:cond_underflow1}), (\ref{eq:cond_total1}) and (\ref{eq:cond_overflow1}) are linear functions of $\mathbf{C}$. Therefore, the optimization problem (\ref{prb:p2}) is convex with respect to $\mathbf{C}$ and we can employ the Lagrangian optimization framework. In particular, we can identify the Karush-Kuhn-Tucker (KKT) conditions and characterize the optimal policy. For this problem, we have the following Lagrangian function using Lagrange multipliers $\mu_t$ and $\lambda_t \geq 0$:

\begin{align}
\textit{L}(\mathbf{\mu}, \mathbf{\lambda}) & = \sum_{j = 1}^{T} \sum_{i = 1}^{M} \left(2^{\frac{C_i(j)}{B_c}}-1\right)\frac{N_0 B_c}{\gamma_i(j)} \nonumber \\
& - \sum_{t=1}^{T-1}\left\{\mu_t \left(\sum_{j = 1}^{t} \sum_{i = 1}^{M} C_i(j) \tau - \sum_{j = 1}^{t} F(j) \right) \right \} \nonumber \\
& - \mu_T \left(\sum_{j = 1}^{T} \sum_{i = 1}^{M} C_i(j) \tau - \sum_{j = 1}^{T} F(j) \right) \nonumber \\
& + \sum_{t=1}^{T} \left\{\lambda_t \left(\sum_{j = 1}^{t} \sum_{i = 1}^{M} C_i(j) \tau - \sum_{j = 1}^{t-1} F(j)-F_{\text{max}} \right) \right\},  \label{eq:obj_lagrangian}
\end{align}
Lagrange multipliers $\mu_t \geq 0, t = 1,2,\ldots,T-1$ and $\mu_T$ are associated with the constraints in (\ref{eq:cond_underflow1}) and (\ref{eq:cond_total1}), respectively. $\lambda_t, t = 1,2,\ldots,T$ are associated with the constraints in (\ref{eq:cond_overflow1}) for all $t$. The additional complimentary slackness conditions are as follows:
\begin{align}
& \mu_t \left(\sum_{j = 1}^{t} \sum_{i = 1}^{M} C_i(j) \tau - \sum_{j = 1}^{t} F(j)\right) = 0, \quad 1 \leq t \leq T-1, \label{eq:cond_underflow_slack}\\
& \mu_T \left(\sum_{j = 1}^{T} \sum_{i = 1}^{M} C_i(j) \tau - \sum_{j = 1}^{T} F(j)\right) = 0, \label{eq:cond_total_slack}\\
& \lambda_t \left(\sum_{j = 1}^{t} \sum_{i = 1}^{M} C_i(j) \tau - \sum_{j = 1}^{t-1} F(j)-F_{\text{max}}\right) = 0, \quad 1 \leq t \leq T. \label{eq:cond_overflow_slack}
\end{align}
Taking the first derivative of (\ref{eq:obj_lagrangian}) with respect to $C_i(j)$, we obtain
\begin{align}
\frac{\partial \textit{L}(\mathbf{\mu}, \mathbf{\lambda})}{C_i(j)} & = 2^{\frac{C_i(j)}{B_c}}\frac{\ln2}{B_c} \frac{N_0 B_c}{\gamma_i(j)} + \left(\tau\sum_{t=j}^{T}(\lambda_t - \mu_t)\right). \label{eq:obj_lagrangian_1st}
\end{align}
From the KKT optimality conditions, the optimal arrival rates $C_i^*(j)$ to the receiver buffer can be obtained after solving
\begin{align}
2^{\frac{C_i^*(j)}{B_c}}\frac{\ln2}{B_c} \frac{N_0 B_c}{\gamma_i(j)} + \left(\tau\sum_{t=j}^{T}(\lambda_t - \mu_t)\right) = 0. \label{eq:derivativeofLagrangian}
\end{align}
Now, using (\ref{eq:derivativeofLagrangian}), the optimal power levels $P_i^*(j)$ in terms of the Lagrange multipliers are expressed as
\begin{align}
P_i^*(j) = & \left(2^{\frac{C_i^*(j)}{B_c}}-1\right)\frac{N_0 B_c}{\gamma_i(j)} \nonumber \\
 = & \frac{\tau B_c}{\ln2} \sum_{t=j}^{T}(\mu_t - \lambda_t) - \frac{N_0 B_c}{\gamma_i(j)} \nonumber \\
 = & \left[W(j) - \frac{N_0 B_c}{\gamma_i(j)}\right]^+,  \label{eq:opt_power}
\end{align}
where $[x]^+ = \max\{0,x\}$, and the water level in time slot $j$ , $W(j)$, is given by
\begin{align}
W(j) = \frac{\tau B_c}{\ln2} \sum_{t=j}^{T}(\mu_t - \lambda_t), \label{eq:water_level}
\end{align}
into which the dual variables (i.e., the Lagrange multipliers) are incorporated.
We have the following characterization regarding the water levels.

\begin{Lem} \label{lem1}
Assume that in the time slot $t = k$, the $k^{th}$ constraint in (\ref{eq:cond_overflow1}) is satisfied with strict inequality (hence the buffer is not full). Then, the optimal water levels at times $t = k$ and $t = k+1$ satisfy $W(k) \geq W(k+1)$. Moreover,  if $\text{Tx}$ transmits a part of the future frames, then $W(k) = W(k+1)$.

Additionally, assume that $F_{\text{max}} > F(t)$ for all $t$ (i.e., the playout buffer size is larger than any frame size in the video sequence, which is required in order to be able to buffer the largest frame). Then, $W(k) < W(k+1)$ occurs only if the $\text{Tx}$ sends a part of the future frames that makes the buffer at the $\text{Rx}$ to be full in time slot $k$.
\end{Lem}

\emph{Proof:} See Appendix \ref{appendix:Theorem1proof}.

One implication of the characterization in Theorem \ref{lem1} is that if $F_{max} = \infty$ (i.e., the playout buffer has infinite size), then all constraints in (\ref{eq:cond_overflow1}) are satisfied with strict inequality and $\lambda_t = 0$ for all $t$ and therefore the optimal water levels $W(j)$ form a monotonically non-increasing sequence i.e., $W(j) \geq W(j+1)$.

In general, it is not an easy task to determine all the water levels $W(j)$ for $1 \leq j \leq T$ by solving (\ref{eq:cond_underflow_slack}), (\ref{eq:cond_total_slack}), (\ref{eq:cond_overflow_slack}) and (\ref{eq:opt_power}). We will employ a dynamic programming based approach to determine the water levels. Assume that $\text{Tx}$ sends only the first $t-1$ frames by time $t-1$ and we have the optimal water levels $W(j)$ up to that time, i.e., for $0 \leq j \leq t-1$. 
Let $W(0) = \infty$ and let $\mathcal{Q}_t = \{q_0, q_1, \ldots, q_{s(t)}\}$ denote the indices of the frames immediately after which the water level becomes different from the previous water level (i.e., a transition occurs in terms of the water levels in the frame $q_k+1$). Equivalently, this also means that the water level $W(q_{k}+1)$ stays the same for frames $q_k+1$ through $q_{k+1}$ (and the water level changes in the next frame with index $q_{k+1}+1$).
Let us set $q_0 = 0$.
Also let $H(j) = \sum_{i=1}^{M} C_i(j)\tau$ denote the data $\text{Rx}$ receives in time slot $j$ for $0 \leq j \leq t$ and $H(0) = 0$.

When a new frame $F(t)$ is added to the video, $\text{Tx}$ sends $F(t)$ in time slot $t$ and the corresponding water level is $W(t) = W_{\text{cur}}$ obtained by solving
\begin{align}
F(t) = \sum_{i=1}^{M} \tau B_c \log\left(1 + \frac{\left[W_{\text{cur}} - \frac{N_0 B_c}{\gamma_i(t)}\right]^+\gamma_i(t)}{N_0 B_c}\right). \label{eq:water_level_temp1}
\end{align}
Now, we compare $W_{\text{cur}}$ with the previous water level $W_{\text{pre}} = W(q_{s(t-1)}+1)$, and consider two cases:
\begin{enumerate}
\item $W_{\text{cur}} \leq W_{\text{pre}}$: By Theorem \ref{lem1}, if the current water level is less than or equal to the previous water level, no further operation or processing is needed and the current water level is the optimal one. We also update $\mathcal{Q}_{t-1}$ to $\mathcal{Q}_{t}$.
%
\item $W_{\text{cur}} > W_{\text{pre}}$: We initialize $k = 0$ and update the new water levels $W(j) = W_{\text{cur}}$ for $q_{s(t-1)-k}+1 \leq j \leq t$ until $W_{\text{cur}} \leq W(q_{s(t-1)-k-1}+1)$ or the buffer is full in time slot $q_{s(t-1)-k}$, by replacing $k$ with $k+1$ and updating $W_{\text{cur}}$ from the following equation:
    \begin{align}
    & F(t) + \sum_{j=q_{s(t-1)-k}}^{t} H(j) \nonumber \\
    = & \sum_{j=q_{s(t-1)-k}}^{t} \sum_{i=1}^{M} \tau B_c \log\left(1 + \frac{\left[W_{\text{cur}} - \frac{N_0 B_c}{\gamma_i(j)}\right]^+\gamma_i(j)}{N_0 B_c}\right). \label{eq:water_level_temp2}
    \end{align}
    Then, the updated power levels $P_i(j)$ and transmitted data $H(j)$ in the corresponding time slots are expressed as follows:
    \begin{align}
    P_i(j) & = \left[W_{\text{cur}} - \frac{N_0 B_c}{\gamma_i(j)}\right]^+, \label{eq:power_update1} \\
    H(j) & = \sum_{i=1}^{M} \tau B_c \log\left(1 + \frac{P_i(j)\gamma_i(j)}{N_0 B_c}\right), \label{eq:data_update1}
    \end{align}
    for $q_{s(t-1)-k}+1 \leq j \leq t$.
    We also update $\mathcal{Q}_{t} = \mathcal{Q}_{t-1}$ after removing all $q_{n}$ with $s(t-1)-k+1 \leq n \leq s(t-1)$ from $\mathcal{Q}_{t-1}$, and $s(t) = s(t-1) - k$.

    Note that if the buffer is not full and the current water level is higher than previous one, $\text{Tx}$ can send the part of current frame $F(t)$ in the previous time slots. And from Theorem \ref{lem1}, the optimal water levels $W(j)$ should be the same for $q_{s(t)}+1 \leq j \leq t$. Since constraints in (\ref{eq:cond_underflow1}) are satisfied at time $t$ and the $\text{Rx}$ receives part of current frame $F(t)$ in previous time slots $q_{s(t-1)-k}+1 \leq j \leq t-1$, constraints in (\ref{eq:cond_overflow1}) are also satisfied at time $t$. However, the buffer might be full at time slot $q_{s(t-1)-k}+1 \leq n \leq t-1$. Therefore, we need to check the overflows from time slots (or equivalently frames) $q_{s(t)}+1$ to $t-1$.
    Initializing $n = q_{s(t)}+1$, we iteratively check if the inequality
    \begin{align}
    \sum_{j = 1}^{n_1} H(j) - \sum_{j = 1}^{n_1-1} F(j) - F_{\text{max}} \leq 0
    \end{align}
    is satisfied or not for any $n \leq n_1 \leq t$ until (\ref{eq:cond_overflow1}) is satisfied at time $t$. If it is satisfied for all $n \leq n_1 \leq t-1$, we get the optimal water levels $W(j)$ and power levels $P_i(j)$ at time $t$. If not, we find the smallest $n_1$, set $f(n_1) = 1$ ($n_1$ is marked as the time that the buffer is full), and update the water levels $W(j) = W_{\text{cur1}}$ for $n \leq j \leq n_1$ and $W(j) = W_{\text{cur2}}$ for $n_1 < j \leq t$ by solving the following equations:
    \begin{align}
    & \sum_{j = 1}^{n_1-1} F(j)+F_{\text{max}} -  \sum_{j = 1}^{n-1} H(j) \nonumber \\
    = & \sum_{j = n }^{n_1} \sum_{i=1}^{M} \tau B_c \log\left(1 + \frac{\left[W_{\text{cur1}} - \frac{N_0 B_c}{\gamma_i(j)}\right]^+\gamma_i(j)}{N_0 B_c}\right), \label{eq:water_level_temp3} \\
    & - \sum_{j = 1}^{n_1-1} F(j) - F_{\text{max}} +  \sum_{j = 1}^{t} H(j) \nonumber \\
    = & \sum_{j = n_1+1 }^{t} \sum_{i=1}^{M} \tau B_c \log\left(1 + \frac{\left[W_{\text{cur2}} - \frac{N_0 B_c}{\gamma_i(j)}\right]^+\gamma_i(j)}{N_0 B_c}\right), \label{eq:water_level_temp4}
    \end{align}
    since the total bits received at $\text{Rx}$ from times $n$ to $t$ is $\sum_{j = n}^{t} H(j)$. Then, the updated power levels and transmitted data in the corresponding time slots are expressed as follows:
    \begin{align}
    P_i(j) & = \left[W_{\text{cur}} - \frac{N_0 B_c}{\gamma_i(j)}\right]^+, \label{eq:power_update2}\\
    H(j) & = \sum_{i=1}^{M} \tau B_c \log\left(1 + \frac{P_i(j)\gamma_i(j)}{N_0 B_c}\right), \label{eq:data_update2}
    \end{align}
    for $n \leq j \leq t$.
    We also update $\mathcal{Q}_{t}$ by adding $n_1$ into it, we also update $s(t) = s(t) + 1$ and $n = n_1 + 1$. Let $f(n_1) = 1$ denote that the buffer storage is full at time $n_1$.
\end{enumerate}

Based on the above detailed descriptions and analysis, the optimal power control algorithm is given below in Algorithm \ref{Alg:Opt_Dynamic}.

\begin{algorithm}[H]
\caption{Dynamic programming based power control algorithm that minimizes the average power consumption}
\begin{algorithmic}[1]\label{Alg:Opt_Dynamic}
\REQUIRE The knowledge of video frame sizes $F(t)$ and CSI $\gamma_i(t)$ for all $t = 1, 2, \ldots, T$. Buffer size $F_{\text{max}}$ at $\text{Rx}$.
\ENSURE The optimal power allocation $\mathbf{P}^*$.
\STATE Initialization: Set $\mathcal{Q}_1 = \{q_{s(0)}\}$, $s(0) = 0$, $q_0 = 0$ and $W(0) = \infty$. $f(j) = 0$ for all $1 \leq j \leq T$.
\FOR {$t = 1:T$}
\STATE Find the current water level $W(t) = W_{\text{cur}}$ by solving (\ref{eq:water_level_temp1}). Initializing the previous water level $W_{\text{pre}} = W(q_{s(t-1)}+1)$. Set $k = 0$.
\WHILE {$W_{\text{cur}} > W_{\text{pre}}$ and $f(q_{s(t-1)-k}) \neq 1$}
    \STATE Update the water levels $W(j) = W_{\text{cur}}$ for $q_{s(t-1)-k}+1 \leq j \leq t$ by solving (\ref{eq:water_level_temp2}).
    \STATE Update corresponding power levels $P_i(j)$ and received amounts of data $H(j)$ for $q_{s(t-1)-k}+1 \leq j \leq t$ by  (\ref{eq:power_update1}) and (\ref{eq:data_update1})
    \STATE Update $k = k+1$ and $W_{\text{pre}} = W(q_{s(t-1)-k}+1)$.
\ENDWHILE
\STATE Remove $q_{s(t-1)-j+1}$ from $\mathcal{Q}_{t-1}$ for all $1 \leq j \leq k$, and set $\mathcal{Q}_{t} = \mathcal{Q}_{t-1}$. Therefore, $s(t) = s(t-1)-k$.
\STATE Initialize $n = q_{s(t)} + 1$, $n_1 = n$.
\WHILE {$n_1 \leq t-1$}
    \IF {$\sum_{j = 1}^{n_1} H(j) - \sum_{j = 1}^{n_1-1} F(j) - F_{\text{max}} > 0$ }
    \STATE Update the water levels $W(j) = W_{\text{cur1}}$ for $n \leq j \leq n_1$ by solving (\ref{eq:water_level_temp3}) and $W(j) = W_{\text{cur2}}$ for $n_1+1 \leq j \leq t$ by solving (\ref{eq:water_level_temp4}).
    \STATE Update corresponding power levels $P_i(j)$ and received amount of data $H(j)$ for $n \leq j \leq t$ by  (\ref{eq:power_update2}) and (\ref{eq:data_update2}).
    \STATE $f(n_1) = 1$, update $\mathcal{Q}_{t}$ by adding $n_1$ to it. Therefore, $s(t) = s(t) + 1$ and $q_{s(t)} = n_1$.
    \STATE Set $n = n_1 + 1$
    \ENDIF
    \STATE $n_1 = n_1 + 1$.
\ENDWHILE
\ENDFOR
\end{algorithmic}
\end{algorithm}

We note that the above power control algorithm is designed for transmission to a single receiver over multiple subchannels. However, this algorithm can also be directly employed when a transmitter sends different video sequences to multiple receivers over different subchannels in an orthogonal fashion.

\subsection{Minimizing the time duration of video streaming}
In the second scenario, the goal is to minimize the duration of time used for transmitting the entire video sequence again under the constraints that $\text{Rx}$ plays the received video without any interruption and missing frames, i.e., without any receiver playout buffer underflows and overflows. Therefore, the optimization problem can be expressed as follows:

\begin{align}
& \min_{\mathbf{P}}T_1 \nonumber \tag{\textbf{P3}}\label{prb:p3} \\
\text{s.t. } & \quad (\ref{eq:cond_underflow1}), (\ref{eq:cond_overflow1}) \nonumber \\
& \sum_{j = 1}^{T_1} \sum_{i = 1}^{M} C_i(j) \tau  = \sum_{j = 1}^{T} F(j), \label{eq:cond_total2}
\end{align}
where constraint (\ref{eq:cond_total2}) describes that $\text{Tx}$ has sent all video data at time $T_1$, and the goal of Problem (\ref{prb:p3}) is to find the minimum $T_1$, which satisfies the constraints (\ref{eq:cond_underflow1}), (\ref{eq:cond_overflow1}) and (\ref{eq:cond_total2}).
Intuitively, minimizing the time consumption implies that $\text{Tx}$ transmits as much video content as possible in each time slot, and hence this minimization problem is equal to maximizing the throughput in each time slot $t$ for $1 \leq t \leq T_1$ until $\text{Tx}$ completes the video transmission assignment at time $T_1$. Also, since the video transmission can potentially be finished very quickly in the absence of any limitations on the transmission power, we impose a maximum power constraint $P_{\text{max}}$ for transmission over $M$ subchannels in each time slot in the optimization problem. The available buffer capacity in time slot $t$ before sending data from the $\text{Tx}$ is expressed as
\begin{align}
A(t) & = F_{\text{max}} - \left(\sum_{j = 1}^{t-1} H(j) - \sum_{j = 1}^{t-1} F(j)\right),  \quad t \geq 2, \nonumber \\
A(1) & = F_{\text{max}},
\end{align}
Thus, the optimization problem (\ref{prb:p3}) is modified as follows:
\begin{align}
& \max_{\mathbf{P}} \sum_{i = 1}^{M} C_i(j) \nonumber \tag{\textbf{P4}}\label{prb:p4} \\
\text{s.t. } & \sum_{i = 1}^{M} P_i(j) \leq P_{\text{max}}, \quad \forall j \geq 1, \label{eq:cond_max_power} \\
& \sum_{i = 1}^{M} C_i(j) \tau \leq R(j) \label{eq:cond_overflow3}
\end{align}
where
\begin{align}
R(j) = \min \left\{ A(j), \quad \sum_{k = 0}^{T} F(k) - \sum_{k = 0}^{j-1} H(k) \right\}
\end{align}
is the minimum value between the available buffer capacity and the remaining video data to be sent in time slot $j$. Thus, (\ref{eq:cond_overflow3}) is the combination of overflow and total video data constraints. In other words, $\text{Tx}$ cannot send an amount of data that is greater than the available buffer capacity or the remaining video bits. We note that Problem (\ref{prb:p4}) does not include the underflow constraint. In this case, underflows are avoided by keeping the maximum power constraint $P_{\text{max}}$ sufficiently large. In particular, in the numerical results in Section \ref{sec:Result}, we set $P_{\text{max}}$ equal to the maximum power required in the solution of the power minimization problem (\ref{prb:p1}) so that we have a fairer comparison between the results of time minimization and power minimization while also avoiding buffer underflows because (\ref{prb:p1}) is formulated to steer clear of any underflows.

We can solve Problem (\ref{prb:p4}) in two steps:
\begin{itemize}
\item First, we ignore the constraint in (\ref{eq:cond_overflow3}). The objective function of Problem (\ref{prb:p4}) is an increasing convex function with respect to $\mathbf{P}$ and the constraint (\ref{eq:cond_max_power}) is linear. Therefore, the optimization problem is a convex optimization problem and it has a unique maximizer. The Lagrangian function for this problem can be expressed as
    \begin{align}
    \textit{G}(\phi) & = \sum_{i = 1}^{M} B_c \log \left(1 + \frac{P_i(j)\gamma_i(j)}{N_0 B_c}\right) \nonumber \\
    & - \phi \left(\sum_{i = 1}^{M} P_i(j)- P_{\text{max}} \right). \label{eq:obj_lagrangian_timemin}
    \end{align}
    By applying the KKT optimality conditions to the Lagrangian function and letting $\frac{\partial \textit{G}(\phi)}{P_i(j)} = 0$, the optimal power levels $P_i^*(j)$ can be expressed in terms of the Lagrange multiplier as follows:
    \begin{align}
    P_i^*(j) = \left[\frac{B_c}{\phi \ln2} - \frac{N_0 B_c}{\gamma_i(j)}\right]^+, \label{eq:opt_power_time_min}
    \end{align}
    where $\phi$ is obtained by solving the following equation:
    \begin{align}
    \sum_{i=1}^{M} \left[\frac{B_c}{\phi \ln2} - \frac{N_0 B_c}{\gamma_i(j)}\right]^+ = P_{\text{max}}. \label{eq:phi_obtain1}
    \end{align}
\item Secondly, we calculate $C_i(j)$ by using the obtained $P_i^*(j)$ in (\ref{eq:opt_power_time_min}). If the obtained power levels $P_i^*(j)$ satisfy the constraint (\ref{eq:cond_overflow3}), $P_i^*(j)$ is the optimal solution. Otherwise, the obtained power levels $P_i^*(j)$ result in buffer overflows. Therefore, the constant power is obtained by solving the following equation:
    \begin{align}
    \sum_{i=1}^{M} B_c \log \left( 1 + \frac{\big[\frac{B_c}{\phi \ln2} - \frac{N_0 B_c}{\gamma_i(j)}\big]^+ \gamma_i(j)}{N_0 B_c}  \right)\tau = R(j). \label{eq:phi_obtain2}
    \end{align}
After obtaining the Lagrange multiplier $\phi$, the optimal power levels are calculated as in (\ref{eq:opt_power_time_min}). The actual throughput is
\begin{align}
H(j) = \sum_{i=1}^{M} B_c \log \left( 1 + \frac{P_i^*(j) \gamma_i(j)}{N_0 B_c}  \right)\tau. \label{eq:throughput}
\end{align}
\end{itemize}
The detailed algorithm is shown below in Algorithm \ref{Alg:Opt_TimeMin}.

\begin{algorithm}[H]
\caption{Power control algorithm for time minimization in video transmission}
\begin{algorithmic}[1]\label{Alg:Opt_TimeMin}
\REQUIRE The knowledge of video frame sizes $F(j)$ and CSI $\gamma_i(j)$ for all $j = 1, 2, \ldots, T$. Buffer size $F_{\text{max}}$ at $\text{Rx}$.
\ENSURE The optimal power allocation $\mathbf{P}^*$ and transmission time $T$.
\STATE Initialization: Set $H(0) = 0$, $t = 0$.

\WHILE {$\sum_{j=0}^{t} H(j) < \sum_{j=1}^{T}F(j)$}
    \STATE Update $t = t+1$.
    \STATE Obtain lagrange multiplier $\phi$ by solving (\ref{eq:phi_obtain1}). After that, optimal power levels $P_i^*(t)$ and throughput $\sum_{i=1}^{M}C_i(j)\tau$ are found.
    \IF {$\sum_{i=1}^{M}C_i(t)\tau > R(t)$}
    \STATE Obtain lagrange multiplier $\phi$ by solving (\ref{eq:phi_obtain2}). After that, optimal power levels $P_i^*(t)$ are found.
    \ENDIF
    \STATE The actual throughput is calculated using (\ref{eq:throughput}).
\ENDWHILE
\STATE The transmission time $T = t$.
\end{algorithmic}
\end{algorithm}

\section{Online Power Control Policies}\label{sec:onlinePolicy}
In the optimal offline policy introduced in the previous subsection, $\text{Tx}$ is assumed to have perfect noncausal CSI for the entire duration of video transmission\footnote{This is a reasonable assumption if the channel conditions vary very slowly and can be predicted accurately.}, and the dynamic programming is employed for solving the optimization problem. In this section, we address online power control policies under the assumption that only the current CSI is available at the $\text{Tx}$ side and future values of channel fading are predicted. We note that online policies are critical for real-time video applications such as live streaming, online gaming, and interactive video. Again, the goal is to lower/minimize the power consumption.

\subsection{The Gauss-Markov Fading and Channel Prediction}
In this section, we introduce a particular channel fading model in order to more concretely address channel prediction. However, the approach and algorithms introduced subsequently can be applied to any channel model and prediction method. The channel is assumed to experience first order Gauss-Markov fading whose dynamics in the $i^{\text{th}}$ subchannel is described by \cite{smisra}
\begin{align}
h_i(j+1) = \alpha h_i(j) + n_i(j+1),  \label{eq:Gauss-Markov}
\end{align}
where $h_i(j)$ is the circularly symmetric complex Gaussian channel fading coefficient at time $j$ with zero mean and variance $\sigma_h^2$. The channel power gain is again denoted as $\gamma_i(j) = |{h_i(j)}|^2$. $n_i(j)$ is the driving noise and $n_i(j) \thicksim \mathcal{CN}(0, (1-\alpha^2)\sigma_h^2)$ where $0 < \alpha < 1$ describes the channel correlation.
Given $h_i(j)$, the predicted channel fading coefficient at time $j+1$ is $\hat{h}_i(j+1) = \alpha h_i(j)$ by using minimum mean square error (MMSE) estimation. Hence, for given the initial fading $h_i(1)$, we have $\hat{h}_i(j+1) = \alpha^j h_i(1)$ for $0 < j < T$. In a video sequence, the number of frames is very large, and $\alpha^j$ becomes very small for a large value of $j$. Due to this, the transmitted video sequence is divided into several groups each with a small number of frames. It is assumed that the group of picture (GoP) size of the video is $N_g$ frames and $L$ GoPs are formed as a group for channel fading coefficient estimation.

We also note that we assume in several numerical results that even the channel correlation may not be perfectly known and the estimated channel correlation coefficient is denoted by $\hat{\alpha}$. In this case, the above prediction formulations above are modified by replacing $\alpha$ with $\hat{\alpha}$.

\subsection{Online power allocation strategy 1 - Grouped water filling (GWF)}
Each group has $N_g L$ frames, and we assume that $\text{Tx}$ knows only the current fading coefficient. For the current time $j$ in group $I$, $\text{Tx}$ predicts the future channel fading coefficients as
\begin{align}
& \hat{h}_i(k + j + (I-1)N_g L) \nonumber \\
= & \alpha^k h(j + (I-1)N_g L),  0 < k \leq N_g L - j, \label{eq:csi_prediction}
\end{align}
which are again the MMSE estimates. The power levels $P_i^*(j + (I-1)N_g L)$ and corresponding received amount of data $H(j + (I-1)N_g L)$ at current time $j$ are obtained by using Algorithm \ref{Alg:Opt_Dynamic} based on the above estimated channel fading coefficients. Following this, we move to the next frame time $j+1$ and the $\text{Tx}$ obtains the perfect knowledge of the current channel fading coefficient and predicts the future channel fading coefficients accordingly. Similarly as in the previous frame time $j$, power levels at time $j+1$ are obtained and the procedure moves to the next frame time until the power levels are obtained for the entire group. In this online algorithm,  the constraints (\ref{eq:cond_underflow_pred}) -- (\ref{eq:cond_overflow_pred}) given on the next page are updated over time.
\begin{figure*}
\begin{align}
& \sum_{k = j}^{l} \sum_{i = 1}^{M} C_i(j+(I-1)N_g L) \tau \geq \max \Big\{\sum_{k = 1}^{l} F(k + (I-1)N_g L) - \sum_{k = 1}^{j-1}H(k + (I-1)N_g L), 0 \Big\}, \forall l = j, \ldots, N_g L - 1, \label{eq:cond_underflow_pred}\\
& \sum_{k = j}^{N_g L} \sum_{i = 1}^{M} C_i(j+(I-1)N_g L) \tau  = \max \Big\{\sum_{k = 1}^{N_g L} F(k + (I-1)N_g L) - \sum_{k = 1}^{j-1}H(k + (I-1)N_g L), 0 \Big\}, \label{eq:cond_total_pred} \\
& \sum_{k = j}^{l} \sum_{i = 1}^{M} C_i(j+(I-1)N_g L) \tau \leq \max \Big \{\sum_{k = 1}^{l-1} F(k + (I-1)N_g L) - \sum_{k = 1}^{j-1}H(k + (I-1)N_g L), 0 \Big \}+F_{\text{max}}, \forall l = j, \ldots, N_g L, \label{eq:cond_overflow_pred}
\end{align}
\end{figure*}

The detailed algorithm is described in Algorithm \ref{Alg:Opt_PowerMin_Prediction} below.
\begin{algorithm}[H]
\caption{Power minimization for video transmission in online fading channel}
\begin{algorithmic}[1]\label{Alg:Opt_PowerMin_Prediction}
\REQUIRE The knowledge of video frame sizes $F(j)$ and channel correlation coefficient $\alpha$. Buffer size $F_{\text{max}}$ at $\text{Rx}$. GoP size $N_g$ and number of GoPs, $L$ in each group. It is assumed that $\frac{T}{N_g L}$ is an integer.
\ENSURE The optimal power allocation $\mathbf{P}^*$.

\FOR {$I = 1:\frac{T}{N_g L}$}
    \FOR {$j = 1:N_g L$}
        \STATE Predict channel fading coefficients $\hat{h}_i(k+(I-1)N_g L)$ by using (\ref{eq:csi_prediction}) for $j < k \leq N_g L$ after perfectly learning the channel fading coefficient $h_i(j+(I-1)N_g L)$ at $\text{Tx}$.
        \STATE Obtain the optimal power level $P_i^*(j + (I-1)N_g L)$ by employing Algorithm \ref{Alg:Opt_Dynamic} based on above predicted channel fading coefficients and calculate received amount of data $H(j + (I-1)N_g L)$.
        \STATE Update the constraints (\ref{eq:cond_underflow_pred}), (\ref{eq:cond_total_pred}) and (\ref{eq:cond_overflow_pred}) for calculation in the next time slot.
    \ENDFOR
\ENDFOR
\end{algorithmic}
\end{algorithm}

\subsection{Online power allocation strategy 2 - Reinforcement Learning}
In this section, the VBR video streaming over a point-to-point link under overflow and underflow constraints is modeled as a Markov decision process (MDP), which provides a suitable mathematical framework for sequential decision making. Following the MDP formulation, we propose a reinforcement learning (RL) algorithm \cite{Sutton}.

As mentioned above, in time slot $t$, $\text{Tx}$ has only causal knowledge about its state. Consequently, since the duration of one time slot, $\tau$, is fixed and known, the selection of $P(t)$ depends solely on the values of the current state, frame size, and current channel fading coefficient at time $t$. Since the selection of $P(t)$ depends only on the current state of the system, the system can be modeled as an MDP. An MDP consists of a set of states $\mathcal{S}$, a set of actions $\mathcal{A}$, a transition model $\mathcal{P}$ and a set of rewards $\mathcal{R}$. At time $t$, the corresponding state $S_t \in \mathcal{S}$ is a function of the stored data (i.e., buffer state) $D(t-1)$ and current channel fading coefficient $h_i(t)$. In our model, the set $\mathcal{S}$, contains an infinite number of possible states since the channel coefficients can take any value in a continuous range. The set of actions $\mathcal{A}$ corresponds to the values of transmit power that can be selected. $\mathcal{A}$ is finite and it is given by $\mathcal{A} = \{P(t), P(t) \in 0 : \delta : P_{\text{max}}\}$ in our model, where $\delta$ is the incremental step size in the power levels. The action dependent transition model defines the transition probabilities denoted as $\mathbb{P}[S_{t+1} \in \mathcal{U} | S_t, P(t)]$, where $\mathcal{U}$ is a measurable subset of $\mathcal{S}$. Finally, the rewards indicate how beneficial the selected $P(t)$ is for the corresponding $S_t$. For each $S_t$ and $P(t)$, we define the reward $R(t) \in \mathcal{R}$ as follows:
\begin{equation}
R(t) = 1 - \frac{P(t)}{P_{\text{max}}}.       \label{eq:reward}
\end{equation}
$R(t)$ can be calculated at the $\text{Tx}$ with the knowledge of $h_i(t)$ and the selected total power $P(t)$. Since $\text{Tx}$ only has information of its state at time $t$, it is preferred to achieve a higher reward at the current $t$ over future ones and the goal is to achieve the highest reward during the entire process. Taking into account this preference, $\gamma \in (0,1]$ is defined as the discount factor of future rewards. The goal is to select $P(t), \forall t$, in order to maximize the expected reward given by
\begin{equation}
R = \lim_{T\rightarrow \infty} \mathbb{E}\left[\sum_{t=1}^T \gamma^t R(t)\right].      \label{eq:exp_reward}
\end{equation}

A policy $\pi$ is defined as a mapping from a given state $S_t$ to the $P(t)$. i.e., $P(t) = \pi(S_t)$. The value functions are defined to measure how good a policy $\pi$ is from $S_t$ onward. These functions can depend solely on the states, called state-value functions or on the state-action pairs, called action-value functions based on different models or applications \cite{Sutton}. The state-value function $V^{\pi}$ is the expected reward given that $\text{Tx}$ follows the policy $\pi$ from state $S_t$ onwards and the action-value function $Q^{\pi}$ is the expected reward starting from the state $S_t$, selecting the action $P(t)$ and following policy $\pi$ thereafter \cite{ortiz}. Following the formulation in \cite{Sutton}, the action-value function is written as
\begin{equation}
Q^{\pi}(S_t, P(t)) = \mathbb{E}\left \{ \sum_{k=0}^{\infty}\gamma^k R(t+k+1) \bigg| S_t, P(t)  \right \}. \label{eq:action_value}
\end{equation}

The optimal policy $\pi^*$ is the policy whose state-value function is greater than or equal to any other policy for every state. The corresponding action-value function for the optimal policy $\pi^*$ is denoted by $Q^*$. Since the value functions can be written in a recursive manner in what is known as the Bellman equations \cite{Sutton}, this recursive representation facilitates the design of RL algorithms \cite{ortiz}. The general form of this Bellman optimality equation for the action-value function is given in \cite{Sutton} as
\begin{align}
& Q^*(S_t, P(t)) =  \nonumber \\
& \sum_{S_{t+1} \in \mathcal{S}} f_{S_t, S_{t+1}}^{P(t)}\left [ R(t) + \gamma \max_{P(t+1)\in \mathcal{A}}Q^*\big(S_{t+1}, P(t+1)\big)  \right ], \label{eq:Bellman}
\end{align}
where $f_{S_t, S_{t+1}}^{P(t)}$ is the transition probability from $S_t$ to $S_{t+1}$ with the corresponding action $P(t)$.

An on-policy temporal difference RL algorithm, termed State-Action-Reward-State-Action (SARSA), is employed in this paper. Since the number of states is infinite, we use a set of binary functions and a linear function approximation to approximate $Q^{\pi}(S_t, P(t))$. The following steps are considered for the implementation of the SARSA RL algorithm. First, the estimation and update of $Q^{\pi}(S_t, P(t))$ is presented. Secondly, the policy for the selection of $P(t)$ according to the estimated $Q^{\pi}(S_t, P(t))$ is defined. Thirdly, the linear function approximation for the computation of $Q^{\pi}(S_t, P(t))$ is applied. Then, the set of binary functions which are used in linear function approximation are linearly combined, and finally, the resulting SARSA algorithm is presented.

\subsubsection{$\epsilon$-greedy policy}
When the number of states is finite, acting greedily with respect to $Q^{\pi}(S_t, P(t))$ leads to the optimal policy \cite{Sutton}. This is because $Q^{\pi}(S_t, P(t))$ is the expected reward given the state-action pair $(S_t, P(t))$ and the action $P(t)$ that maximizes $Q^{\pi}(S_t, P(t))$ leads to the highest expected reward. However, it has no opportunity to explore transmit power values that can potentially lead to higher rewards if $\text{Tx}$ always acts greedily. In order to solve this problem, the $\epsilon$-greedy policy is considered instead:
\begin{equation}
\text{Pr}\left[ P(t) = \max_{p \in \mathcal{A}} Q^{\pi}(S_t, p)  \right] = 1 - \epsilon, \quad  0 < \epsilon < 1.
\end{equation}
In another words, with probability $\epsilon$, $\text{Tx}$ selects a transmit power value from the action set $\mathcal{A}$ randomly. Since the chosen action $P(t)$ may lead to buffer overflow or underflows, we can precalculate $P_{\text{min}}(t)$ and $P_{\text{max}}(t)$, which denote the minimum and maximum transmission power levels that satisfy buffer overflow and underflow constraints by letting $P_{\text{min}}(t) \leq P(t) \leq \min \{P_{\text{max}}(t), P_{\text{max}}\}$. If $P_{\text{min}}(t) > P_{\text{max}}$, the underflow occurs and cannot be avoided and we let $P(t) = P_{\text{max}}$. However, we can always choose a lower power level to avoid the occurrence  of an overflow.

\subsubsection{Linear function approximation}
We employ the on-policy SARSA algorithm in this paper due to its favorable convergence properties when linear function approximation is used \cite{Sutton}. In SARSA, the next state-action pair $(S_{t+1}, P(t+1))$ is obtained from the current state-action pair $(S_{t}, P(t))$ with a given policy $\pi$, and $Q^{\pi}(S_t, P(t))$ is estimated from this transition process. When the system is in state $S_t$, $\text{Tx}$ selects $P(t)$ following policy $\pi$. After that, it obtains a reward $R(t)$ and moves to state $S_{t+1}$. According to the current values of $Q^{\pi}(S_t, P(t))$ and the policy $\pi$, the next action $P(t+1)$ is selected. After that, action value $Q^{\pi}(S_t, P(t))$ is updated using the previous experience and the current value. The updating rule in the SARSA algorithm is given as follows:
\begin{align}
& Q^{\pi}(S_t, P(t)) \leftarrow  \nonumber \\
& Q^{\pi}(S_t, P(t))(1 - \beta_t) + \beta_t [R(t) + \gamma Q^{\pi}(S_{t+1}, P(t+1))], \label{eq:value_update}
\end{align}
where $\beta_t$ is a small positive fraction which influences the learning rate.

In order to handle the infinite number of states, the concept of linear function approximation is considered \cite{ortiz}. With linear function approximation, $Q^{\pi}(S_t, P(t))$ is represented by a linear combination of $K$ feature functions $\text{f}_k(S_t, P(t))$, $k = 1, 2, \ldots, K$. Each $\text{f}_k(S_t, P(t))$ maps the state-action pair $(S_t, P(t))$ into a feature value. Let $\mathbf{f} \in \mathbb{R}^{K \time 1}$ be a vector containing the feature values for a given state-action pair and let $\mathbf{w} \in \mathbb{R}^{K \time 1}$ be the vector containing the corresponding weights indicating the contribution of each feature to the value. Therefore, the action-value function approximation is given as \cite{Sutton}
\begin{equation}
\hat{Q}^{\pi}(S_t, P(t), \mathbf{w}) = \mathbf{f}^{\text{T}}\mathbf{w}.      \label{eq:value_approx}
\end{equation}

In approximate SARSA, the action-value updates are performed on the weights instead of in (\ref{eq:value_update}). At time $t$, the vector $\mathbf{w}$ is updated in the direction that reduces the error between $Q^{\pi}(S_t, P(t))$ and $\hat{Q}^{\pi}(S_t, P(t), \mathbf{w}) $ following the gradient descent approach. The update rule is expressed as
\begin{align}
\mathbf{w} = \mathbf{w} + \alpha_t \bigg[ & R(t) + \gamma \hat{Q}^{\pi}\big(S_{t+1}, P(t+1), \mathbf{w}\big) \nonumber \\
& - \hat{Q}^{\pi}\big(S_t, P(t), \mathbf{w}\big) \bigg] \triangledown_{\mathbf{w}}\hat{Q}^{\pi}(S_t, P(t), \mathbf{w}), \label{eq:weight_update}
\end{align}
where $\triangledown_{\mathbf{w}}\hat{Q}^{\pi}(S_t, P(t), \mathbf{w})$ is the gradient of $\hat{Q}^{\pi}(S_t, P(t), \mathbf{w})$ with respect to $\mathbf{w}$, and
\begin{equation}
\triangledown_{\mathbf{w}}\hat{Q}^{\pi}(S_t, P(t), \mathbf{w}) = \mathbf{f}.     \label{eq:weights_grad}
\end{equation}

\subsubsection{Feature functions}
The definition of the feature functions is an important step in the implementation of the approximate SARSA algorithm. The features should provide a good model of the effect of possible transmit power values on the state. In our scenario, the most important characteristics are the capacity of the playout buffer and the minimum required video data to be played at $\text{Rx}$. $K = 3$ binary functions are used by taking into consideration playout buffer size and the power allocation problem.

Since overflows are undesirable, the first feature function $\text{f}_1(S_t, P(t))$ indicates if a given $P(t)$ avoids the overflow of the data in the playout buffer at $\text{Rx}$. Additionally, it evaluates if the given action $P(t)$ fulfills the constraint in (\ref{eq:overflow1}). The function is assigned value ``$1$" if no overflow is caused, and is ``$0$" otherwise. Now, the corresponding feature function is written as
\begin{equation}
\text{f}_1(S_t, P(t)) = \begin{cases}
    1, & D(t) \leq F_{\text{max}}\\
    0, & \text{otherwise}
\end{cases}.
\end{equation}

The second feature considers the underflow event. Since $\text{Rx}$ needs to play the $t^{\text{th}}$ frame at time $t$, the amount of stored date in the playout buffer at time $t$ should be no less than the $t^{\text{th}}$ frame size in order to avoid an underflow. Similarly, the second feature function is assigned value ``$1$" if no underflow occurs and the corresponding feature function is formulated as
\begin{equation}
\text{f}_2(S_t, P(t)) = \begin{cases}
    1, & D(t) \geq F(t)\\
    0, & \text{otherwise}
\end{cases}.
\end{equation}

The third feature function $\text{f}_3(S_t, P(t))$ addresses the power allocation problem. We have determined in the offline case that a directional water-filling algorithm can be used to optimally allocate the power. However, the knowledge of future channel coefficients is unavailable in the online scenario. Therefore, we propose to use past channel realizations to estimate the mean value of the distribution of the channel gain and to perform water-filling considering the estimated mean value of the channel gain and the current channel realization. For the estimation, the sample mean estimator is used and the estimated mean value $|\hat{h}_i(t)|^2$ is calculated as
\begin{equation}
|\hat{h}_i(t)|^2 = \frac{1}{t}\sum_{j=1}^{t}|h_i(t)|^2.
\end{equation}

The reason for applying water-filling between $|\hat{h}_i(t)|^2$ and $|h_i(t)|^2$ is that we are assuming that $|\hat{h}_i(t)|^2$ approximates the state of the channel in the subsequent time slot, and consequently the amount of data required has to be considered. And the value of this amount is
\begin{equation}
D_n(t) = \max \{0, F(t)+F(t+1)-D(t)\}.
\end{equation}

The water level $v(t)$ is the solution of
\begin{align}
& \sum_{i=1}^{M}\Bigg[\log_2\left\{1+ \left[v(t) - \frac{N_0 Bc}{|h_i(t)|^2}\right]^+\frac{|h_i(t)|^2}{N_0 B_c}\right\} \nonumber \\
& +  \log_2\left\{1+ \left[v(t) - \frac{N_0 Bc}{|\hat{h}_i(t)|^2}\right]^+\frac{|\hat{h}_i(t)|^2}{N_0 B_c}\right\}   \Bigg] = \frac{D_n(t)}{\tau B_c},
\end{align}
where $[x]^+$ is the maximum value between $x$ and $0$.

The power allocated to the $i^{\text{th}}$ subchannel and the total power are given by
\begin{align}
& p_{i,\text{WF}}(t) = \max \left\{0,  v(t) - \frac{N_0 Bc}{|h_i(t)|^2}\right\}, \\
& p_{\text{WF}}(t) = \sum_{i = 1}^{M}p_{i,\text{WF}}(t),
\end{align}
respectively.
Since power levels are assumed to have discrete values, the calculated $p_{\text{WF}}(t)$ has to be rounded such that $p_{\text{WF}}(t) \in \mathcal{A}$ holds. $\text{f}_3(S_t, P(t))$ is now expressed as
\begin{equation}
\text{f}_3(S_t, P(t)) = \begin{cases}
    1, & \delta \lfloor \frac{p_{\text{WF}}(t)}{\delta} \rfloor = P(t)\\
    0, & \text{otherwise}
\end{cases}
\end{equation}
where $\delta$ is the step size and $\lfloor x \rfloor$ is the rounding operation to the nearest integer less than or equal to $x$.

\subsubsection{Approximate SARSA}
The detailed approximate SARSA algorithm for power control in VBR video wireless transmission system is shown in Algorithm \ref{Alg:Opt_SARSA}. It has been shown in \cite{Gordon} that if $\beta_t$ satisfies $\sum_{t}\beta_t = \infty$ and $\sum_{t}\beta_t^2 < \infty$ and the policy is not changed during the learning process, the approximate SARSA algorithm converges to a bounded region with probability one. $\beta_t = \frac{1}{t}$ is assumed in our scenario.

\begin{algorithm}[H]
\caption{Approximate SARSA for power control}
\begin{algorithmic}[1]\label{Alg:Opt_SARSA}
\REQUIRE The knowledge of video frame sizes $F(t)$ and current CSI $h_i(t)$. Buffer size $F_{\text{max}}$ at $\text{Rx}$.
\ENSURE The optimal power allocation $\mathbf{P}^*$.
\STATE Initialization: Initialize $\gamma$, $\beta_1$, $\epsilon$ and $\mathbf{w}$.
\STATE Observe $S_t$
\STATE Select $P(t)$ using $\epsilon$-greedy
\FOR {$t = 1:T$}
\STATE Transmit using the selected power $P(t)$.
\STATE Calculate corresponding reward $R(t)$ by using (\ref{eq:reward}).
\STATE Observe next state $S_{t+1}$
\STATE Select next transmit power $P(t+1)$ using $\epsilon$-greedy
\STATE Update $\mathbf{w}$ by using (\ref{eq:weight_update}).
\ENDFOR
\end{algorithmic}
\end{algorithm}

\section{Numerical Results}\label{sec:Result}
To evaluate the performance of the proposed power control and video transmission strategies in the simulations, we have used VBR video traces \textit{Tokyo Olympics}, \textit{NBC News} and \textit{Terminator} in all the simulations from the Video Trace Library hosted at Arizona State University \cite{reisslein}. The video parameters are listed in Table \ref{Tab:video_parameters}. The playout buffer size is set to be $1.5$ times the largest frame size among the frames to be transmitted. $P_{\text{max}}$ in time minimization (TM) scheme is set to the maximum power level allocated among all frame time slots in the power minimization (PM) scheme.

We further assume that the bandwidth of each subchannel is $B_c = 10$ kHz and the number of subchannels is set to $M = 100$. Therefore the total bandwidth for the system is $1$ MHz\footnote{We note that if videos (e.g., HD or 4K) with higher resolutions than the ones described in Table \ref{Tab:video_parameters} are used, then bandwidth levels of more than $1$ MHz would be needed to support the larger throughput required by these videos while the general characterizations we have provided in the numerical results would not be significantly altered.}.

\begin{table}[h]
\begin{center}

\caption{Parameters of the video sequences} \label{Tab:video_parameters}
    \begin{tabular}{| c | c |}
    \hline
    Resolution         & $352 \times 288$   \\ \hline
    FPS & $30$  \\ \hline
    Encoder & JSVM(9.15) \\ \hline
    GoP pattern & G16B3 \\ \hline
    Layer & 2 \\ \hline
    \end{tabular}
\end{center}
\end{table}

\subsection{Offline power control}
In the offline power strategy, we assume Rayleigh fading channels in the simulations, for which the normalized path gain is exponentially distributed with probability density function $f(\gamma_{i}) = \exp\{\frac{-\gamma_{i}}{G_{i}}\}/G_{i}$ where path gain averages are $G_{i} = 2$ for subchannels, where $i \in \{1, 2, \ldots, M\}$.

\begin{figure}
\centering
\includegraphics[width=0.48\textwidth]{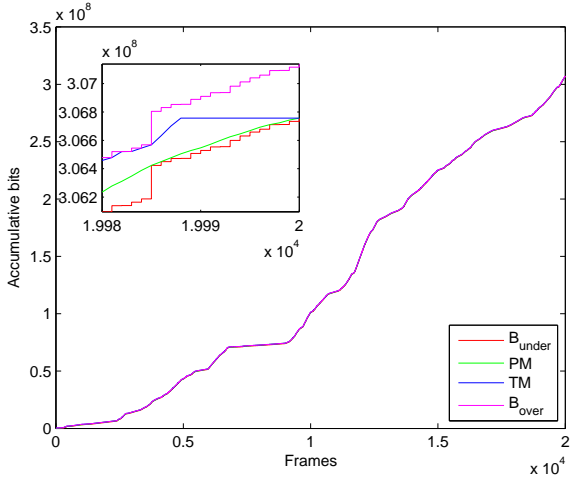}
\caption{\small{Cumulative transmitted data in 2 different schemes.}}\label{fig:Cumu_Tokyo}
\end{figure}

\begin{figure}
\centering
\begin{subfigure}[b]{0.48\textwidth}
\centering
\includegraphics[width=\textwidth]{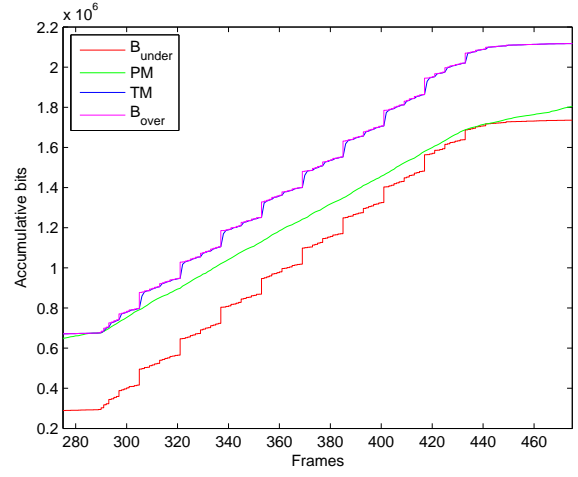}
\caption{Curves between frame time slot $275$ and $475$}\label{fig:Tokyo_Segments:Cumu_Segment1}
\end{subfigure}
\begin{subfigure}[b]{0.48\textwidth}
\centering
\includegraphics[width=\textwidth]{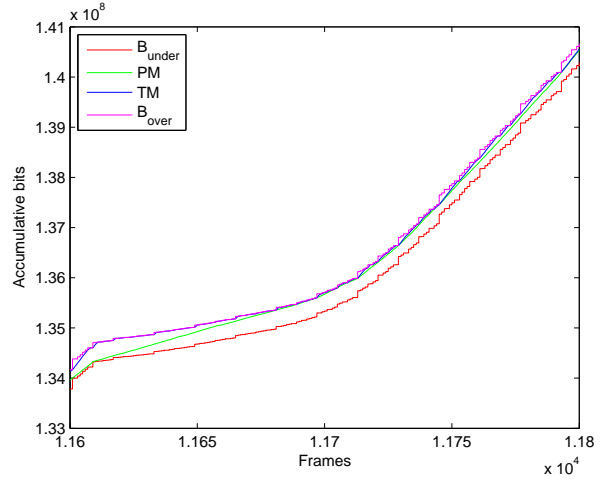}
\caption{Curves between frame time slot $11600$ and $11800$}\label{fig:Tokyo_Segments:Cumu_Segment2}
\end{subfigure}
\caption{\small{The cumulative overflow, transmission, and consumption curves when transmitting \textit{Tokyo} in two different time periods (a) frame time slot $275$-$475$; (b) frame time slot $11600$-$11800$.}}\label{fig:Tokyo_Segments}
\end{figure}

Fig. \ref{fig:Cumu_Tokyo} shows the consumption curves of the buffer at $\text{Rx}$ from frame-time slot $1$ to $20000$. The cumulative overflow, transmission, and consumption curves for TM and PM schemes are plotted when transmitting  \textit{Tokyo Olympics}. The higher slope of the underflow curve means that frame sizes during that time period are larger, and hence the bit rates should correspondingly be larger as well. PM scheme completes video transmission mission at the end of frame time slot $20000$ while TM scheme finishes it at frame time slot $19988$. This saving in time depends on the buffer size and the maximum transmission power. When the transmission power is large enough, larger buffer size leads to more saving in time. In Fig. \ref{fig:Tokyo_Segments}, we observe that both cumulative consumption curves obtained by considering PM and TM schemes are in between the underflow curve and the overflow curve, implying that $\text{Rx}$ plays the video smoothly without any interruptions or missing frames. The consumption curve of TM scheme is always above that of PM due to the fact that TM scheme attempts to send as much data as possible in each frame time slot during the entire video transmission session, and consumption curve of TM reaches the overflow curve in most of the frame time slots. In such cases, the bit rate with TM is in general larger than that with the PM scheme as expected since time minimization requires the maximization of the rate per frame while being cognizant of the buffer overflows. In Fig. \ref{fig:Tokyo_Segments:Cumu_Segment1}, consumption curve of PM reaches the overflow curve at frame time $280$ and then decreases to the underflow curve at frame time $430$ with a lower bit rate. The reason is that, the frame sizes after frame time $430$ are small enough, which leads the buffer to store enough frames for playing. Parts of further future frames are not needed to be stored in order to save power. However, in Fig. \ref{fig:Tokyo_Segments:Cumu_Segment2}, consumption curve of PM reaches the underflow curve at frame time $11610$ and then increases to the overflow curve at frame time $11680$ because the frame sizes after frame time $11680$ are very large and the buffer has to store enough frames for playing the video without any interruption and lowering the power consumed after frame time $11680$.

\begin{figure}
\centering
\begin{subfigure}[b]{0.48\textwidth}
\centering
\includegraphics[width=\textwidth]{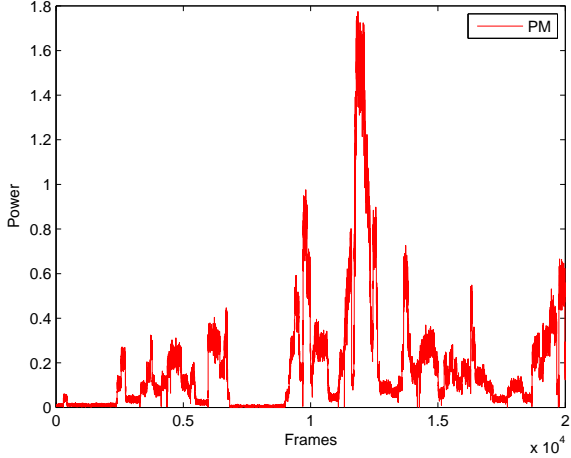}
\caption{Power consumption in the PM scheme}\label{fig:Tokyo_Powers:Power_PM}
\end{subfigure}
\begin{subfigure}[b]{0.48\textwidth}
\centering
\includegraphics[width=\textwidth]{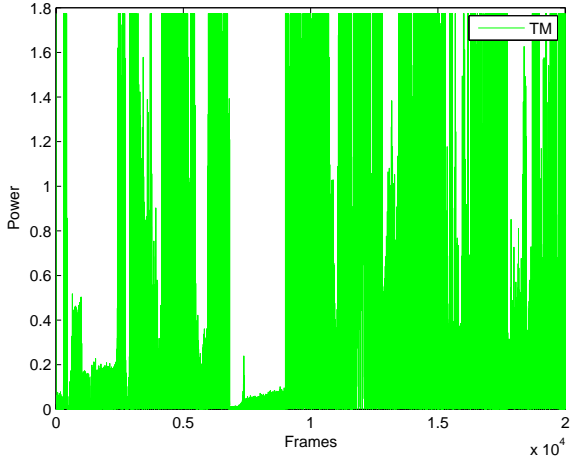}
\caption{Power consumption in the TM scheme}\label{fig:Tokyo_Powers:Power_TM}
\end{subfigure}
\caption{\small{The power consumption when transmitting \textit{Tokyo} in two different schemes (a) power minimization (PM); (b) time minimization (TM).}}\label{fig:Tokyo_Powers}
\end{figure}

Fig. \ref{fig:Tokyo_Powers} displays the consumed power in each frame time slot during the entire video transmission in PM and TM schemes. Fig. \ref{fig:Tokyo_Powers:Power_PM} demonstrates that power levels around time slot $12000$ are the highest since the slope around that time is the largest as seen in Fig. \ref{fig:Cumu_Tokyo}, meaning that the frame sizes around that time slot are the largest and $\text{Tx}$ needs much more energy for completing the transmission of such large-sized frames. There also exists several peaks, which are located at time slots with larger frame sizes compared to other time slots. Fig. \ref{fig:Tokyo_Powers:Power_TM} shows that the peak transmission power level in the TM scheme is around $1.8$ Watts and $\text{Tx}$ transmits frames by using $P_{\text{max}}$ most of the time because the buffer at $\text{Rx}$ tries to store as much data as possible in each time slot without violating the buffer overflow and maximum transmission power constraints. And the buffer is full after receiving data from $\text{Tx}$ if the transmission power level is less than $P_{\text{max}}$ in this time slot.
Otherwise, the buffer can store more data by using higher power level. The average power levels are $0.1827$ and $0.2635$ Watts in PM scheme and TM scheme, respectively.

Table \ref{Tab:power} shows the power consumptions for transmitting different video sequences. The number of frames is $20000$ for all video sequences. The power is in the units of Watts. We notice that PM scheme saves much power (\%30, \%37, and \%48 power savings, respectively, in \emph{Tokyo Olympics, NBC News} and \emph{Terminator} videos) while TM scheme saves only a small number of time slots in video transmission. If the buffer size is larger, the saving in time can be more.
\begin{table}[h]
\begin{center}
\caption{Power consumption for different video sequences} \label{Tab:power}
    \begin{tabular}{| c | c | c | c | c |}
    \hline
                     & $P_{\text{max}}$ & PM & TM & Time saving (slots)  \\ \hline
    \textit{Tokyo Olympics} & $1.7745$ & $0.1827$ & $0.2635$ & $12$  \\ \hline
    \textit{NBC News} & $4.4814$ & $0.6382$ & $1.0240$ & $10$ \\ \hline
    \textit{Terminator} & $4.3549$ & $0.2939$ & $0.5670$ & $13$ \\ \hline
    \end{tabular}
\end{center}
\end{table}

\begin{figure}
\centering
\includegraphics[width=0.48\textwidth]{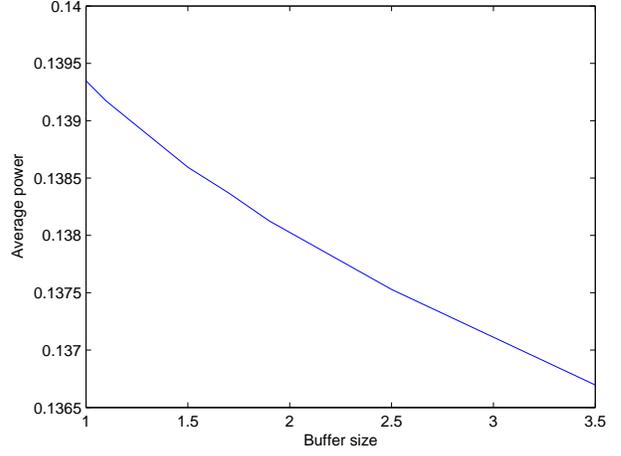}
\caption{\small{The relation between buffer size at $\text{Rx}$ and average power level at $\text{Tx}$.}}\label{fig:Power_Buffer}
\end{figure}

Changing buffer size $F_{\text{max}}$ also affects the power consumption at $\text{Tx}$. Fig. \ref{fig:Power_Buffer} shows the relation between buffer size at $\text{Rx}$ and average power level at $\text{Tx}$. The average power level at $\text{Tx}$ decreases as the buffer size at $\text{Rx}$ increases since $\text{Rx}$ can store more data before it is played, and the instantaneous powers can be adjusted more efficiently. If the buffer size increases from $1.5$ to $2.5$ times the largest frame size, the average power level drops from $0.1386$ W to $0.1375$ W. Note that even small power saving can translate into substantial savings in energy especially if the video sequence is long since average energy will be average power times the duration of the video.

\subsection{Online power control}
For online transmission strategies, we assume Gauss-Markov Rayleigh fading channels in the simulations.
The path gain average is again $G_{i} = 2$ for subchannels, where $i \in \{1, 2, \ldots, M\}$. VBR video trace \textit{Tokyo Olympics} is used in the simulations. GoP is $N_g = 16$ frames and channel correlation coefficient is $\alpha = 0.99$ unless specified otherwise.

\begin{figure}
\centering
\includegraphics[width=0.48\textwidth]{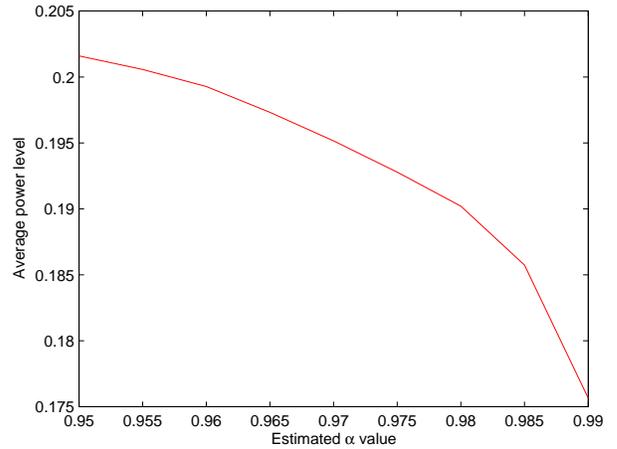}
\caption{\small{Average power level with different $\hat{\alpha}$ values.}}\label{fig:Power_vs_Alpha}
\end{figure}

Fig. \ref{fig:Power_vs_Alpha} demonstrates the relation between average power level and estimated channel correlation coefficient $\hat{\alpha}$ value with group size, $L = 4$. We observe that $\text{Tx}$ sends video sequence with lower average power level as $\hat{\alpha}$ approaches to the true $\alpha$ value of $0.99$. (\ref{eq:csi_prediction}) indicates that smaller $\hat{\alpha}$ leads to larger channel fading coefficient difference between two time slots and $\hat{h}_i(j)$ is lower than the case with high $\hat{\alpha}$ value. Thus, the first several time slots in a group need to send more data by using higher power level in the case of small $\hat{\alpha}$ value compared to the case of large $\hat{\alpha}$ value. In another words, the frames in one group are sent just within first few number of slots when the value of $\hat{\alpha}$ is small, and are sent using all the available frame time slots when $\hat{\alpha}$ has a larger value. Note also that, the estimation quality improves when $\hat{\alpha}$ increases from $0.95$ to $0.99$ with step size $0.05$. Therefore, $\text{Tx}$ sends video sequences with higher average power level if $\hat{\alpha}$ has a smaller value.

\begin{figure}
\centering
\begin{subfigure}[b]{0.48\textwidth}
\centering
\includegraphics[width=\textwidth]{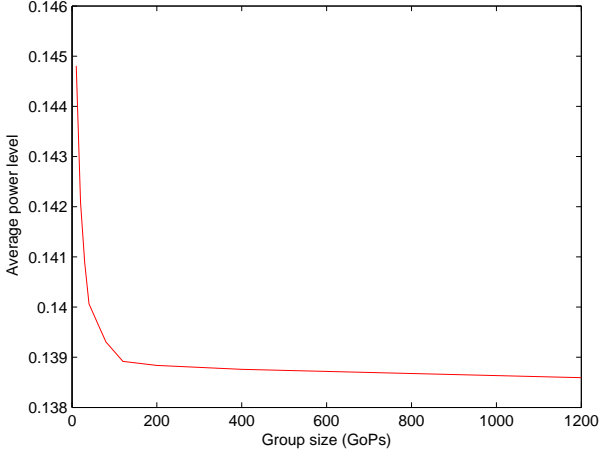}
\caption{Average power level with offline power control}\label{fig:Tokyo_Powers:Power_vs_GoPs_Perfect}
\end{subfigure}
\begin{subfigure}[b]{0.48\textwidth}
\centering
\includegraphics[width=\textwidth]{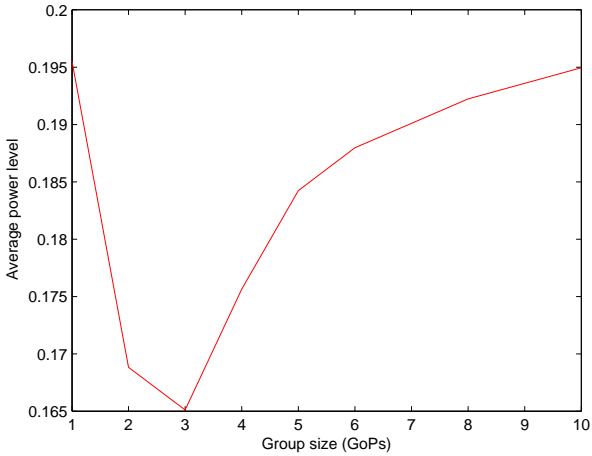}
\caption{Average power level with online power control}\label{fig:Tokyo_Powers:Power_vs_GoPs_Predict}
\end{subfigure}
\caption{\small{The average power level when transmitting \textit{Tokyo} in (a) offline; (b) online strategies with different group sizes.}}\label{fig:Tokyo_Powers:Power_vs_GoPs}
\end{figure}

Fig. \ref{fig:Tokyo_Powers:Power_vs_GoPs} demonstrates the relation between average power level and group size $L$ ($L$ GoPs) considering offline and online power control strategies. Fig. \ref{fig:Tokyo_Powers:Power_vs_GoPs_Perfect} demonstrates that the average transmission power gets smaller when group size $L$ increases in the case of offline power control, because the system minimizes power consumption in each individual group. In another words, the number of groups is larger if the group size $L$ is smaller. Thus, the power minimization strategy is implemented as a unit to a larger number of frames if the group size is larger. This further leads to lower power levels. However, in online power control, average transmission power initially decreases and then starts getting larger as the group size $L$ grows further as shown in Fig. \ref{fig:Tokyo_Powers:Power_vs_GoPs_Predict}. At first, the communication system consumes more power if the group size is small since the online strategy is implemented in each individual group with small number of frames. Theoretically, if the strategy is implemented in a group with larger number of frames, the average power is lower. However, as discussed in the case of varying $\hat{\alpha}$ values, the larger group size leads to smaller channel coefficients among the latter frame time slots in each group. Thus, the entire group of frames need to be sent in the first few frame time slots and the system consumes more power. Therefore, the average power level eventually starts increasing when the group size $L$ grows beyond a threshold.

Next, we address the performance achieved with the SARSA algorithm. VBR video trace \textit{Terminator} is used in simulations. GoP is $N_g = 16$ frames, $L = 4$ and channel correlation coefficient $\alpha$ varies. We now assume that the Tx perfectly knows the values of $\alpha$ (i.e., we have $\hat{\alpha} = \alpha$). For each $\alpha$ value, we generate the channel side information $10$ times, and for each set of channel side information, we run the algorithm $10$ times. Therefore, for each $\alpha$ value, we run the code $100$ times.

\begin{figure}
\centering
\includegraphics[width=0.48\textwidth]{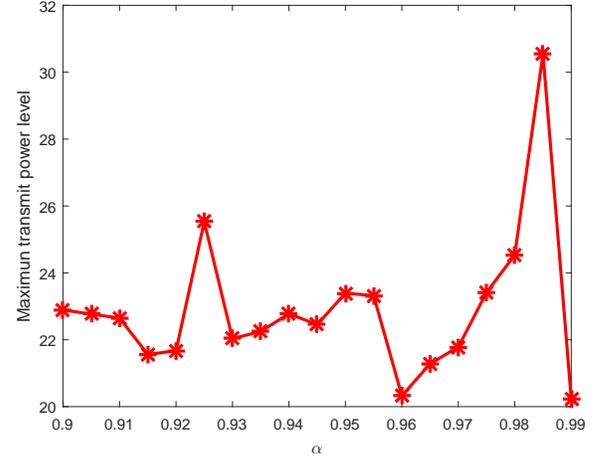}
\caption{\small{Maximum transmit power levels in GWF method.}}\label{fig:MaxPower_alpha}
\end{figure}

Fig. \ref{fig:MaxPower_alpha} shows the maximum transmit power levels used in the GWF algorithm in order to avoid underflow and overflows. Note that since channel correlation varies, the maximum transmit power levels changes depending on $\alpha$ values.

\begin{figure}
\centering
\includegraphics[width=0.48\textwidth]{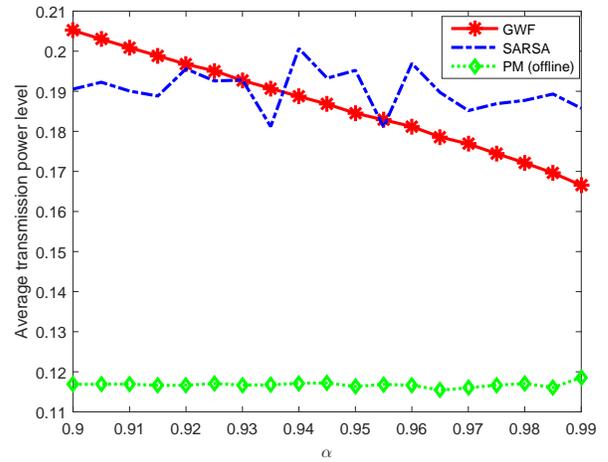}
\caption{\small{Transmit power levels for different $\alpha$ values.}}\label{fig:Power_alpha}
\end{figure}
Fig. \ref{fig:Power_alpha} plots the average transmit power levels for different $\alpha$ values attained with grouped water-filling (GWF) and reinforcement learning SARSA strategies as well as the optimal offline PM algorithm. Note that in the offline policy, all channel fading coefficients are assumed to be known non-causally prior to video transmission. Consequently, the offline PM algorithm attains the lowest average transmission power levels.  On the other hand, for the online policies, we see that if the channels are more correlated, meaning that the value of $\alpha$ is larger and the future channel side information can be estimated more accurately, the average transmit power level tends to be lower. We also observe that, with SARSA strategy, the average power levels fluctuate but within a certain small range. Note that SARSA scheme estimates the channel side information just in the next time slot. Fig. \ref{fig:Power_alpha} also shows that the transmit power attained with the GWF strategy is higher than that achieved with the SARSA strategy when $\alpha$ is smaller than a certain value, because small $\alpha$ value leads to a large channel estimation error that propagates over the entire group of frames used in the GWF strategy. SARSA strategy just estimates the channel in the next time slot, and thus the error is smaller than in the GWF strategy. Correspondingly, SARSA reinforcement learning performs better for smaller values of $\alpha$. However, if channel is highly correlated (implying a high $\alpha$ value), the GWF strategy is close to the optimal solution in each group of frames. Therefore, GWF strategy consumes less power than the SARSA strategy.

\begin{figure}
\centering
\includegraphics[width=0.48\textwidth]{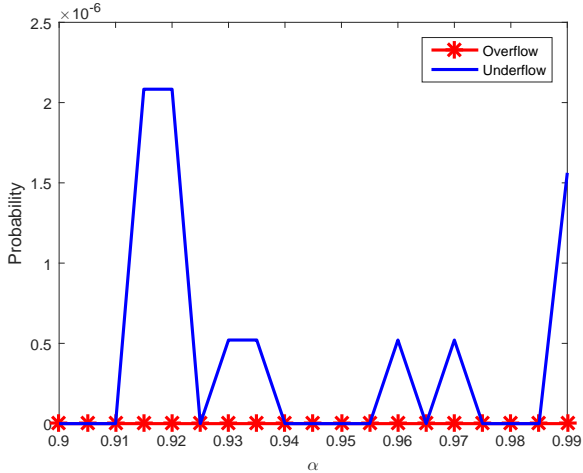}
\caption{\small{Underflow probability and overflow probability for different $\alpha$ values.}}\label{fig:Probability_alpha}
\end{figure}
Fig. \ref{fig:Probability_alpha} presents the underflow probability and overflow probability for different $\alpha$ values in SARSA strategy\footnote{We note that the underflow probability is equal to the rebuffering ratio (defined as the rebuffering duration divided by the entire video playback duration) in our simulation results due to the fact that underflow probability is computed as the ratio of the number of time slots in which underflow/rebuffering has occurred over the total number of time slots.}. Since the GWF is the approach to find the optimal power with predicted channel information while avoiding underflows and overflows, the underflow and overflow probabilities are $0$. In SARSA strategy, the overflow probability is $0$ as we noted in the discussion of the $\epsilon$-greedy policy. On the other hand, if the channel conditions are very poor and the maximum transmit power $P_{\text{max}}$ cannot support the minimum amount of video data to be sent to the $\text{Rx}$, underflow event happens. However, the underflow probability is very small as shown in Fig. \ref{fig:Probability_alpha}.

\section{Conclusion}\label{sec:Conclusion}
In this paper, we have studied both offline and online power control strategies for wireless VBR video streaming over multiple subchannels. We have addressed power control at the $\text{Tx}$ subject to VBR video characteristics and playout buffer underflow and overflow requirements to satisfy the quality of experience (QoE) expectations of the users. We have identified directional water filling as the optimal offline power control policy and developed algorithms considering both power minimization (PM) and time minimization (TM). We have shown that the PM strategy can lead to more than 30\% savings in consumed energy compared to the TM strategy.
Following the analysis of the optimal offline policy, the algorithm is modified to solve the extended optimization problem in the online setting. This first online policy is called the grouped water-filling (GWF). As a second online strategy, the reinforcement learning based SARSA algorithm is proposed to determine an efficient power allocation policy and the results have shown that the RL SARSA performs better than the GWF strategy if the channel is not highly correlated. Overall, we have developed efficient and dynamic resource allocation strategies for wireless VBR video streaming by identifying the optimal offline power control policies and proposing novel online power control schemes based on GWF and reinforcement learning.

\appendix

\subsection{Proof of Theorem \ref{lem1}} \label{appendix:Theorem1proof}
If the $k^{th}$ constraint in (\ref{eq:cond_overflow1}) is satisfied with strict inequality, i.e., $\sum_{j = 1}^{k} \sum_{i = 1}^{M} C_i(j) \tau < \sum_{j = 1}^{k-1} F(j)+F_{\text{max}}$, then we have $\lambda_k = 0$ by slackness conditions in (\ref{eq:cond_overflow_slack}). And from (\ref{eq:water_level}), we have
\begin{gather}
W(k) = W(k+1) + \frac{\tau B_c}{\ln2} (\mu_k - \lambda_k). \label{eq:WkWk+1}
\end{gather}
Since $\mu_k \geq 0$, we readily observe that $W(k) \geq W(k+1)$.

Moreover, if in time slot $k$, $\text{Tx}$ sends a part of the future frames, then the $k^{\text{th}}$ constraint in (\ref{eq:cond_underflow1}) is satisfied with strict inequality. This means that in that case we have $\mu_k = 0$ by slackness conditions in (\ref{eq:cond_underflow_slack}). Hence, by (\ref{eq:WkWk+1}), $W(k) = W(k+1)$.

When $F_{\text{max}}$ is  greater than any frame size in the video sequence, i.e., $F_{\text{max}} > F(t)$ for all $t$, the constraints in (\ref{eq:cond_underflow1}) and  (\ref{eq:cond_overflow1}) at $t = k$ cannot be satisfied with equality at the same time. In other words, we cannot have $\lambda_k > 0$ and $\mu_k > 0$ simultaneously in the same time slot $k$. From (\ref{eq:WkWk+1}), $W(k) < W(k+1)$ implies that $\mu_k - \lambda_k < 0$. Therefore, $W(k) < W(k+1)$ occurs only if $\mu_k = 0$ and $\lambda_k > 0$. $\lambda_k > 0$ means that the constraint in (\ref{eq:cond_overflow1}) at $t = k$ is satisfied with equality. Thus, we have $W(k) < W(k+1)$ only if $\text{Rx}$ has received a part of the future frames that makes the buffer to be full in time slot $k$. \hfill $\blacksquare$

\bibliographystyle{IEEEtran}
\bibliography{VBR_Journal}
\end{document}